\documentclass[preprint,12pt]{elsarticle}
\usepackage{bm}

 \usepackage{graphics}

\usepackage{amssymb}
\usepackage{amsmath}
\usepackage{color}

\biboptions{comma,square}

\journal{Physica A}

\begin{document}

\begin{frontmatter}



\title{Boundary conditions and heat resistance
at the moving solid--liquid interface}


\author[omsk]{G.\,L.~Buchbinder \corref{pkg}}
\ead{glbuchb@yahoo.com, glb@omsu.ru}
\author[uni-jena,urfu]{P.\,K. Galenko}
\cortext[pkg]{Corresponding author}
 \ead{peter.galenko@uni-jena.de}

\address[omsk]{Department of Physics, Omsk State University, peace Avenue 55A,644077 Omsk, Russia}
\address[uni-jena]{Friedrich-Schiller-Universit{\"{a}}t Jena,
Physikalisch-Astronomische Fakult{\"{a}}t, D-07743 Jena, Germany}
\address[urfu]{Ural Federal University,
Laboratory of Multi-Scale Mathematical Modeling,
620002 Ekaterinburg, Russla}


\begin{abstract}
Boundary conditions for the  solid-liquid interface of the
solidifying  pure melt  have been derived. In the derivation the
model of Gibbs interface is used. The boundary conditions  include
both  the state quantities  of bulk phases are taken at the
interface and  the quantities characterizing interfacial surface
such as the surface temperature and  the surface heat flux.
Introduction of the surface temperature as an independent variable
allows us to describe the scattering energy at the interface.  For
the steady-state motion of the planar interface the expression for
the temperature discontinuity across the phase boundary has been
obtained. Effect of Kapitza resistance on the interface velocity is
considered. It is shown that heat resistance  leads to non-linearity
in solidification kinetics, namely, in ``velocity-undercooling''
relationship. The conditions  of the steady--state motion of the planar
interface has been found.
\end{abstract}

\begin{keyword}
Kapitza resistance\sep Interface \sep Model \sep Crystal



\end{keyword}

\end{frontmatter}


\newpage

\section{Introduction}\label{Intro}

Thermal processes at the phase interface essentially influence
crystal growth and ultimately define  final structure  of
solidifying materials. Well known examples are dendritic
growth~\cite{JL80} and the heat trapping with the temperature jump
at the interface due to absence of local equilibrium both
 at the interface ~\cite {U07} and the bulk phases ~\cite{GD00}.
 The temperature jump has been first detected at the interface
between liquid helium and metallic substrate and it is well known as
the Kapitza resistance against heat transport~\cite{PLK41}.

The existence of heat resistance means that the energy exchange at
the interface includes not only the transverse energy transfer but,
generally speaking, the energy flux along the interface.  This has
been namely observed in experiments on water droplet evaporation
from the warmed Au-substrate~\cite{GW11}.  It  has been found that a
big part of heat energy is transferred parallel to the water-metal
boundary through the adsorbed layer, leading to measurable
interfacial temperature discontinuity.

Taking into account this experimental results, Palmieri et
al.~\cite{PWD12} have supposed that the temperature discontinuity
can also be essential at the moving phase interface, especially
during fast solidification. These authors have considered influence
of Kapitza resistance on the planar front dynamics during of
solidification from undercooled state. In their analysis, the kinetic
equation  for the rate  of phase changes  has been used that
determines in fact the interface temperature for the case of the
temperature discontinuity across the interface. For the system where
all material constants are identical within  both phases the
interface temperature was introduced as a half-sum of boundary
temperatures of the each phase.

In the linear approximation with respect to an equilibrium boundary
conditions at the interface were obtained in Ref.~\cite{PWD12} and
present the set of Onsager relations. In these relations, the
dissipative currents are the solidification rate and the energy flux
through the interface and driving forces are
interphase temperature jump and departure of the interfacial
temperature from the equilibrium temperature $T_M$ of
solidification. It should be noted that the notion about the
interfacial boundary conditions as Onsager relations have also been
suggested in the work of Brener and Temkin~\cite{BT12} (see also
Ref.~\cite{CN80}) although without introduction of the interface
temperature as independent variable and with a different choice of
thermodynamic fluxes and their conjugated driving forces.

A general formalism of the derivation  of boundary conditions at the
moving phase surface has been developed by  Bedeaux, Albano и Mazur
(BAM)~\cite{BAM76,ZB82} (see also Ref.~\cite{W67}). The
thermo-hydrodynamical variables  for a surface have been introduced
in the  spirit of Gibbs concept of the dividing surface for which
the extensive surface variables are determined as the surface
excesses of the corresponding bulk quantities~\cite{KB08}.

The BAM--formalism~\cite{BAM76,ZB82} was applied by Caroli et al. to
the solidification  problem ~\cite{CCR84}. They have derived the
boundary conditions at interface for a binary system  but neglected
by all surface contributions. In particular case of
the pure melt and without convection in bulk phases the boundary
conditions of Caroli et al.~\cite{CCR84} coincide with those ones
suggested by Brener and Temkin~\cite{BT12}.

In the present work the general approach based on BAM-formalism~\cite{BAM76,ZB82} is
applied to the derivation of boundary conditions at the interfacial
surface of the solidifying pure liquids taking into account the
contributions from surface variables including the heat flux
along the surface (longitudinal heat flux). The obtained boundary conditions are then
applied to the analysis of heat resistance effect on planar  front
dynamics in solidification of a stagnant liquid.

The work is organized as follows. In Section~\ref{sec:SerfVar}, the
variables characterizing the interface are introduced and the
expression for surface entropy production, obtained in \ref{app1},
is given. Section~\ref{sec:boundCond} is devoted to the derivation
of the boundary conditions which then  applied to the steady state
motion of the interface in Section~\ref{sec:Steady}. In Section~\ref{sec:Tjamp},
the temperature  jump through the interface is
calculated and various regimes of the flat front motion are
analyzed depending on the thermal Kapitza resistance coefficient.
Conclusion and discussion of the main results  are given in Section
\ref{sec:Concl}. In~\ref{app1}, the BAM-method is summarized in the
context of solidifying pure liquid. Finally, in \ref{app2},
the modified kinetic equation for the rate of phase change
derived in the framework of the Gibbs model of an interface.

\section{Surface wariables and entropy production}
\label{sec:SerfVar}

At equilibrium, the Gibbs's capillary model considers interface as a
surface of a zero thickness (i.e., a sharp interface) separating
phases and which is situated within the diffuse zone. The
surface is characterizes by finite densities of extensive thermodynamic
quantities are defined as surface excess densities of bulk state
quantities ~\cite{KB08}). To describe non-equilibrium states at
the moving interface, the surface densities of
dissipative fluxes have also been introduced~\cite{BAM76,W67}. According to
BAM-formalism~\cite{BAM76}, if $x$ the density per unit volume of some extensive
state quantity (mass density, velocity, energy fluxes and so on),
then at any space-time point $x$ can be represented as
\begin{equation}\label{1}
x({ \bf r},t) = x^+({ \bf r},t){\mathit\Theta }^+(f) + x^s({ \bf
r},t){\mathit\delta }^s({ \bf r},t) + x^-({ \bf r},t){\mathit\Theta
}^-(f),
\end{equation}
where the equation $f({\bf r},t) = 0$ defines the phase interface,
 $f({\bf r},t) > 0 $ defines the region occupied by liquid,
 and $f({\bf r},t) < 0 $ defines the solid phase, $\mathit\Theta ^\pm (f)$ are
Heviside functions
\[\mathit\Theta ^+(f) = \left\{
\begin{array}
{rll}
1,  &f > 0\\
0,  &f\leqslant0
\end{array}\right .
\hspace{1cm}
 \mathit\Theta ^-(f) = \left\{
\begin{array}
{rll}
0,  &f \geqslant 0 \\
1,  &f<0
\end{array}\right .,
\]
$x^\pm $ are the densities of bulk variable~$x$ for each phase,
which are labeled as $+$ for the liquid and as $-$ for the solid,
$x^s$ is the surface density. The surface $\delta$-function  is
defined as
\[{\mathit\delta }^s({ \bf r}t) =|\nabla f|\delta(f({\bf r}t)),\]
where $\delta(f)$ is the standard $\delta$-function.

Definition~(\ref{1}) means that the  densities~$x$ become singular
at the interface. Quantities $x^s$  vary only along the surface but do
not normally to the surface. Therefore, the normal derivations of
these quantities are  zero.

In accordance to Eq.~(\ref{1}), the densities of internal energy $u$,
 entropy~$s$, as well as the entropy production $\sigma$ can be written
 in the form
\begin{eqnarray}
&&u = u^+{\mathit\Theta }^+ + u^s{\mathit\delta }^s + u^-{\mathit\Theta }^-, \label{2}\\
&&s = s^+{\mathit\Theta }^+ + s^s{\mathit\delta }^s + s^-{\mathit\Theta }^-, \label{3}\\
&& \sigma = \sigma^+{\mathit\Theta }^+ + \sigma^s{\mathit\delta }^s
+ \sigma^-{\mathit\Theta }^-\label{4},
\end{eqnarray}
where $u^s$, $\sigma^s$, and $s^s$ are the surface densities of
these quantities and $s^\pm$, $\sigma^\pm$, and $u^\pm$ are their
bulk densities. All quantities in Eqs.~(\ref{2})-(\ref{4}) are the
functions of space-time point, however, the surface densities are
defined only on the interface.

In a simplest approximation, we shall assume that the mass densities
of both phases are constants. In addition, the dividing surface can
be chosen in such a manner that one of excess densities may take
zero value~\cite{KB08}.   The mass density $\varrho^s$ is usually
chosen as  such density.  Also, we assume that the flow in phases is
absent such that the velocity field becomes:
$$\bm{\upsilon} = \left\{
\begin{array}
{ll}
0,  &f > 0 \\
\bm{\upsilon}^s, &f = 0 \\
0,& f < 0
\end{array}\right .,$$
where~${\upsilon}^s$ is the velocity of the boundary (the front velocity)
which is changing along the surface.

In the Gibbs model, the equilibrium surface temperature $T^s$ is defined
by the equality~\cite{KB08}
\begin{equation}\label{5}
ds^s = \frac{1}{T^s}du^s\quad  \text{with}~~ \varrho^s = 0.
\end{equation}
In equilibrium, one has $T^s = T^+ = T^-$, where  $T^\pm$ are the
temperatures of bulk phases. In the absence of equilibrium, the
equation (\ref{5}) defines the temperature  of the  surface element
having the internal energy $u^s$ and the entropy $s^s$ \cite{BAM76}.
Equation (\ref{5}) means that there is local equilibrium along
surface. It should be noted that the hypothesis of local equilibrium
with respect to the system volume  means that any physically small
volume is characterized  by some  unique temperature $T$. In this
sense any physically small volume including an element of the
interfacial surface is not in the local thermodynamic equilibrium
because such  volume is characterized by the three  temperatures
$T^- \neq T^s \neq T^+ $, where $T^\pm$ are the limiting  values of
the bulk temperatures  at the interface   from its both sides in
phases.

As it  has been noted in the introductory section, Onsager relations at the interface can be considered  as
boundary conditions. The derivation of these relations is based on the
principle of positive definiteness of the entropy~\cite{GM69}. The
detailed derivation for the surface entropy production $\sigma^s$
for the solidification problem  is provided in \ref{app1} and it
gives
\begin{eqnarray}
 &&\sigma^s =  -
 {\bf J}^s_q \Bigl(\frac{1}{T^s}\Big)^2\nabla_{||} T^s
 + \Bigl[\frac{\mu\varrho }{T}\Bigr]_-\upsilon^s_{n} +
  [J_{qn} - h\upsilon^s_{n}]_+ \Bigl(\frac{1}{T^+} - \frac{1}{T^-}
  \Bigr)\nonumber \\
  &&\phantom{aaaa}+ [J_{qn} - h\upsilon^s_{n}]_{-} \Bigl[ \frac{1}{2}\Bigl( \frac{1}{T^+} + \frac{1}{T^-}\Bigr) -
  \frac{1}{T^s}\Bigr]\,,\label{6}
\end{eqnarray}
where ${\bf J}^s_q $ is the  surface density of the  heat flux ${\bf
J}_q$ defined in consistency with Eq.~(\ref{1}). In what follows,
all bulk quantities, entering in the equations at the interface
[see, for example, Eq.~(\ref{6})], present their limiting values
$x^\pm$  at the interface, and, for the jump of $x$ through the interface,
the following notations have been introduced
\[[x]_- = x^+ - x^-,\]
\[[x]_+ = \frac{1}{2}(x^+ + x^-)\,.\]
Therefore, in Eq.~(\ref{6}), $\mu$ is the chemical potential,
$\varrho$ is the mass density, $h$ is the enthalpy and $T^\pm$ are
the boundary  temperatures of phases. The heat flux in the normal
direction to the interface is denoted as $J^\pm_{qn} = {\bf
n}\cdot{\bf J}^\pm_q$, where ${\bf n}$ is the unit vector to the
interface and $v^s_n = {\bf n}\cdot{\bm \upsilon}^s$ is the normal
component of the interface velocity. The surface heat flux ${\bf
J}^s_q$ lies in the tangent plane and its scalar product with any
vector along normal is  zero. The symbol $||$ defines the component
of a vector along the tangent plane  to the interface.

\section{Boundary conditions}
\label{sec:boundCond}

Equality~(\ref{6}) can be represented as a sum of products of the
dissipative fluxes and conjugated driving forces. Then, taking into
account the positivity of entropy production, $\sigma^s \geqslant 0$,
one can write down linear  relations  between this quantities. We
take $\nabla_{||}T^s$, $[\mu\varrho/T]_-$, $[1/T]_-$, and
\[
 \frac{1}{2}\Bigl( \frac{1}{T^+} + \frac{1}{T^-}\Bigr) -
  \frac{1}{T^s},\]
as the driving forces and ${\bf J}^s_q $, $\upsilon^s_{n}$, and
$[J_{qn} - h\upsilon^s_{n}]_+$ , $[J_{qn} - h\upsilon^s_{n}]_-$ as
the dissipative fluxes. Thus, there are the vectorial force-flux pair
and three scalar force-flux pairs in Eq.~(\ref{6}). Taking into
account tensorial nature of the various  quantities with their isotropy
at the surface and in bulk, we have linear laws as
\begin{equation}
{\bf J}^s_q  = - \lambda^s\nabla_{||} T^s,  \label{7}
\end{equation}
and
\begin{eqnarray}
  &&\upsilon^s_{n} = L_{00}\Bigl[\frac{\mu\varrho}{T}\Bigr]_-  + L_{01}\Bigl(\frac{1}{T^+} - \frac{1}{T^-}
  \Bigr) +  L_{02}\Bigl[ \frac{1}{2}\Bigl(
\frac{1}{T^+} + \frac{1}{T^-}\Bigr) -
  \frac{1}{T^s}\Bigr],  \label{8}\\
& &[J_{qn} - h\upsilon^s_{n}]_+ = L_{10}
\Bigl[\frac{\mu\varrho}{T}\Bigr]_-  + L_{11}\Bigl(\frac{1}{T^+} -
\frac{1}{T^-} \Bigr) + \nonumber\\
&&\phantom{aaaaaaaaaaaaaaaaaaaaaaaaaaaa}  + L_{12}\Bigl[
\frac{1}{2}\Bigl( \frac{1}{T^+} + \frac{1}{T^-}\Bigr) -
  \frac{1}{T^s}\Bigr],  \label{9}\\
&&[J_{qn} - h\upsilon^s_{n}]_- = L_{20}
\Bigl[\frac{\mu\varrho}{T}\Bigr]_-  + L_{21}\Bigl(\frac{1}{T^+} -
\frac{1}{T^-} \Bigr) + \nonumber\\
&& \phantom{aaaaaaaaaaaaaaaaaaaaaaaaaaaa} +  L_{22}\Bigl[
\frac{1}{2}\Bigl( \frac{1}{T^+} + \frac{1}{T^-}\Bigr) -
  \frac{1}{T^s}\Bigr],  \label{10}
\end{eqnarray}
where $\lambda^s$  has a meaning of surface heat conductivity and
$L_{ij}$ are the kinetic coefficients satisfying the Onsager reciprocal conditions,
$L_{ij} = L_{ji}$.

Expression~(\ref{7}) connects the surface heat flux ${\bf J}^s_q$
with the temperature $T^s$ analogously to the Fourier law for the
bulk transport. Relations~(\ref{8})-(\ref{10}) give boundary
conditions for the bulk quantities. Expression~(\ref{8}) defines the
interface velocity. Equality~(\ref{10}) presents a difference
between the energy fluxes on the both sides of the interface (energy
flux jump)  and Eq.~(\ref{9}) defines their half-sum. As it follows
from Eq.~(\ref{8}), thermodynamic conditions for crystallization
(i.e., conditions of non-zero interface velocity) are determined by
the difference of the chemical potentials, the temperature jump, and
also by the difference of boundary  temperatures  in phases from the
interface temperature $T^s$.

In addition, relations~(\ref{8})-(\ref{10}) should be completed
for the interface temperature $T^s$ (see Eq.~(\ref{A21}))
\begin{equation}\label{11}
c^s\frac{d^sT^s}{dt} +  \,\mbox{div}\, {\bf J}^s_q  +
T^ss^s\mbox{div}\,{\bm{\upsilon}}^s  =  [h\upsilon^{s}_n -
    J_{qn}]_-\,,
\end{equation}
where $c^s$ is the interfacial heat capacity. The  dissipation of an
energy and its transfer along the interface give rise to the
interface temperature changes and are taken into account in
Eq.~(\ref{11}) by the energy flux jump in the  right hand side of
this equation  and the surface heat flux ${\bf J}^s_q$.  At the
steady-state solidification (i.e., in the regime with the constant
interface velocity along the interface) and for homogeneous
interface (${\bf J}^s_q = 0$), the left hand side of Eq.~(\ref{11})
is equal to zero that leads to the conservation of the energy flux
through the interface.

System of equations~(\ref{7})-(\ref{10}) and (\ref{11}) represents
complete system of boundary conditions at the phase interface. After
exclusion of the flux~(\ref{7}), these boundary conditions can be
considered as the system of four equations with four unknown
quantities: two boundary temperatures~$T^\pm$, growth
velocity~$\upsilon_n^s$ and interface temperature~$T^s$. In what
follows, we consider application of this system of equations to the
motion of planar solidification front with constant velocity.

\section{Steady-state interface motion}{\label{sec:Steady}

Consider the homogeneous interface which is moving with the constant
velocity and  the energy dissipation at the interface is absent. In
this case Eq.~(\ref{11}) gives equality of energy fluxes at the both
sides of interface:
\begin{equation}\label{13}
[J_{qn}- h\upsilon^s_n]_- = 0.
\end{equation}
Then, from Eq.~(\ref{10}) one gets the equality
\begin{eqnarray*}
&&L_{20} \Bigl[\frac{\mu\varrho}{T}\Bigr]_-  +
L_{21}\Bigl(\frac{1}{T^+} - \frac{1}{T^-}
  \Bigr) +  L_{22}\Bigl[ \frac{1}{2}\Bigl(
\frac{1}{T^+} + \frac{1}{T^-}\Bigr) -
  \frac{1}{T^s}\Bigr] = 0\,,
\end{eqnarray*}
which defines the surface temperature as
\begin{eqnarray}
  &&  \frac{1}{T^s} = \frac{1}{2}\Bigl(
\frac{1}{T^+} + \frac{1}{T^-}\Bigr)  + \frac{L_{20}}{L_{22}}
\Bigl[\frac{\mu\varrho}{T}\Bigr]_- +
\frac{L_{21}}{L_{22}}\Bigl(\frac{1}{T^+} - \frac{1}{T^-}
  \Bigr).
  \label{14}
\end{eqnarray}
With the accuracy up to small thermodynamic forces,
$[\mu\varrho/T]_-$ ,$[1/T]_- $, Eq.~(\ref{14}) gives
\begin{equation}\label{15}
    \frac{1}{T^s} = \frac{1}{2}\Bigl(
\frac{1}{T^+} + \frac{1}{T^-}\Bigr)
\end{equation}
As it follows now from Eq.~(\ref{8}), the interface velocity is
defined by two quantities: $[\mu\varrho/T]_-$ and $[1/T]_- $.
Because $\upsilon^s_n$ must be zero at $T^s = T^+ = T^- = T_M$, then
$[\mu\varrho/T]_-$ should take zero value for the same conditions.
In this case, one can write the following expansion
\begin{equation}\label{16}
\Bigl[\frac{\mu\varrho}{T}\Bigr]_- = {\cal A}\Bigl(\frac{1}{T^s} -
\frac{1}{T_M}
  \Bigr) + {\cal B}\Bigl(\frac{1}{T^+} - \frac{1}{T^-}
  \Bigr) + \cdots\,,
\end{equation}
where $ {\cal A}$ are $ {\cal B}$ some constants. Substituting first
two terms of Eq.~(\ref{16}) into Eqs.~(\ref{8})-(\ref{10}), taking
into account Eq.~(\ref{15}), and re-denoting kinetic coefficients
($L \rightarrow L'$), one gets
\begin{eqnarray}
  &&\upsilon^s_{n} = L_{00}'\Bigl(\frac{1}{T^s} - \frac{1}{T_M}
  \Bigr) + L_{01}'\Bigl(\frac{1}{T^+} - \frac{1}{T^-}
  \Bigr), \label{17}\\
&& [J_{qn} - h\upsilon^s_{n}]_+ = L_{10}'\Bigl(\frac{1}{T^s} -
\frac{1}{T_M}
  \Bigr) + L_{11}'\Bigl(\frac{1}{T^+} - \frac{1}{T^-}
  \Bigr),   \label{18}\\
&&[J_{qn}]_- = L^s{\upsilon}^s_n + c_p[T]_-{\upsilon}^s_n.
\label{19}
\end{eqnarray}
Due to equality~(\ref{13}), the left hand side of Eq.~(\ref{18})
defines the energy flux $J_E$ through the phase interface.
Note that the coefficients $ L'_{ij}$ do not already satisfy
the reciprocal Onsager relations, i.e. $L'_{ij} \neq L'_{ji}$.

Equation~(\ref{19}) presents modified boundary condition of Stefan
and it can be obtained from the equality~(\ref{13}) rewritten as
\begin{equation}\label{20}
J^+_{qn} - J^-_{qn} =  (h^+ - h^-){\upsilon}^s_n\,,
\end{equation}
where $h^+ - h^- = h^+(T^+) - h^-(T^-)$, and
\begin{eqnarray}
  h^\pm(T^\pm) &=& h^\pm(T^s + T^\pm - T^s)\nonumber\\
\phantom{aaaaa}&\thickapprox & h^\pm(T^s) +
  \Bigl(\frac{\partial h^\pm}{\partial T^\pm}\Bigr)_{T^s}(T^\pm -
  T^s)+\cdots ,\label{21}
\end{eqnarray}
The derivative of enthalpy with respect to temperature defines the
heat capacity at the constant pressure. Taking the heat capacities in both phases to be constant,
$\Bigl(\partial h^\pm/\partial T^\pm\Bigr)_{T^s} =
c_p$, one gets
\begin{eqnarray}
   && h^+ - h^- = h^+(T^s) - h^-(T^s) + c_p (T^+ - T^-) \nonumber\\
&&\phantom{h^+ - h^-  }= L^s + c_p(T^+ - T^-)\,.
\label{22}
\end{eqnarray}
Here $L^s = h^+(T^s) - h^-(T^s)$ can be interpreted as latent heat
of solidification.  Finally,  substitution of Eq.~(\ref{22}) into
Eq.~(\ref{20}) gives Eq.~(\ref{19}).

At the end of this section, one has to note that, considering the
interface velocity $\upsilon^s_{n}$ as a flux allows us compare
Eq.~(\ref{17}) with the expression for the front velocity obtained
from kinetic equation (see Appendix B and Section~\ref{sec:Tjamp}).

\section{The temperature jump}
\label{sec:Tjamp}

The obtained boundary conditions are not limited by a  specific
model of heat transport in the bulk system and have a general
meaning. Using Eqs.~(\ref{17})-(\ref{19}), we now calculate the
temperature jump $\Delta T = T^+ - T^-$ at the phase interface and
the front velocity in a framework of parabolic model of heat
transport.

Let us consider  now crystallization of undercooled liquid. Assume
that  the planar interface  is moving  with the constant velocity
$v_n^s = V$ in the direction of their normal being z-axis. In this
case  the temperature and the heat flux in the system volume satisfy
the one-dimension diffusion equation
\begin{eqnarray}
  \frac{\partial T^\pm(zt)}{\partial t} &=& \varkappa\,\frac{\partial^2 T^\pm(zt)}{\partial z^2} \label{23}\\
 \frac{\partial J_q^\pm(zt)}{\partial t} &=& \varkappa\,\frac{\partial^2 J_q^\pm(zt)}{\partial z^2}\,,  \label{24}
\end{eqnarray}
where $\varkappa$ is the heat conductivity which is assumed to be
equal in both phases. Further we assume  the heat diffusion in solid
phase is absent. Then  Eqs.~(\ref{23})-(\ref{24}) have solutions in
a form of traveling waves. In the liquid region, $Z = z - Vt > 0$,
one has
\begin{eqnarray}
&&T^+(zt) =  T_0 + A_1 e^{-VZ/\varkappa}, \label{25} \\
&& J_q^+(zt)  =  A_2 e^{-VZ/\varkappa},     \label{26}
\end{eqnarray}
and in the region of solid phase, $Z < 0$, one gets
\begin{equation}\label{27}
T^-(zt) =   A_3\,, \hspace{0.5cm} J_q^-(zt)  =  0\,,
\end{equation}
where $T_0$ is the initial temperature of liquid and $A_i$ are
constants. Then, at the interface, $Z = 0$, one can get the following
values of temperature and heat flux for both phases (further the
bulk state quantities at the interface will be wrote without
arguments $z t$)
\begin{equation}\label{28}
T^+ =  T_0 + A_1\,,\hspace{0.5cm}
 J_q^+ \equiv J^+_{qn}  = c_pV A_1\,,
\end{equation}
where we have used the equality $A_2 = c_pVA_1 $ following from the
energy conservation law,   $c_p\,\partial T/\partial t = -\partial
J_q/\partial z$, and
\begin{equation}\label{29}
T^-=   A_3\,,\hspace{0.5cm} J_q^- \equiv J^-_{qn} =  0.
\end{equation}

Thus, we have three the independent constant $ A_1$, $ A_3$ and $V$
in Eqs.~(\ref{25})-(\ref{27}) which can be defined from three
equations (\ref{17})-(\ref{19}).  To determinate  these quantities
we can rewrite Eqs.~(\ref{17})-(\ref{19}) at small values of the deviation
$T^\pm - T_M $ using the following approximations
\begin{eqnarray}
  &&\frac{1}{T^s} - \frac{1}{T_M} = \frac{1}{2}\Bigl(\frac{1}{T^+} + \frac{1}{T^-}\Bigr) - \frac{1}{T_M} \nonumber\\
&& \phantom{aaaaaaaaa} \simeq \frac{1}{2T_M^2}(T^+ + T^- - 2T_M)\label{30}\\
  && \frac{1}{T^+} - \frac{1}{T^-} \simeq  \frac{1}{T_M^2}(T^- - T^+)\label{31}\,.
\end{eqnarray}
Further, taking into account Eq.~(\ref{21}), one can write
\begin{eqnarray}
   &&[h]_+ = \frac{1}{2}(h^+ + h^-) = \frac{1}{2}[h^+(T^s) +
   h^-(T^s)]\nonumber\\
   &&\phantom{aaaaa} + \frac{1}{2}c_p(T^+ + T^- - 2T^s) + \cdots \nonumber\\
   &&\phantom{aaaaaaaaaaaaaa}\simeq \frac{H^s}{2} + \mathcal{O}(\Delta T^2)\,,\label{32}
\end{eqnarray}
where $H^s/2 = [h^+(T^s) + h^-(T^s)]/2$ can be interpreted as the interfacial enthalpy.

As a result, instead of Eqs.~(\ref{17})-(\ref{19}), one finds
\begin{eqnarray}
  &&V = \frac{L_1}{2}(T^+ + T^- - 2T_M) + L_2(T^- - T^+),   \label{33}\\
&&J_E = [J_{qn}]_+  - \frac{1}{2}H^s V \nonumber\\
&&\phantom{aa} = \frac{L_3}{2}(T^+ + T^- - 2T_M) + L_4(T^- - T^+),  \label{34}\\
&&[J_{qn}]_- = L^sV + c_pV(T^+ - T^-),\label{35}
\end{eqnarray}
where the following notations
\[L_1 = \frac{L'_{00}}{T_M^2},\quad L_2 = \frac{L'_{01}}{T_M^2},\quad L_3 = \frac{L'_{10}}{T_M^2},\quad L_4 = \frac{L'_{11}}{T^2_M}\,,\]
have been introduced. Note that the coefficient \[R = L^{-1}_4\] has
a meaning of the Kapitza coefficient.

Expressions for the temperature jump and the boundary temperatures
can now be obtained from Eqs.~(\ref{28})-(\ref{29}) and
Eqs.~(\ref{34})-(\ref{35}) as
\begin{eqnarray}\label{36}
  && \Delta T = R\frac{Vh^- + L_3(T_0 + L^s/c_p -T_M)}{1 - R(L_3 -
   c_pV)/2},\\
  && T^+ = T_0 + L^s/c_p + \Delta T,    \label{37}\\
&& T^- =  T_0 + L^s/c_p,    \label{38}
\end{eqnarray}
where $h^- = h^-(T^s)$. To simplify  further calculations we assume
approximately  for the evaluation the enthalpy that
\begin{equation}
    h^- = h^-(T^s) \approx c_pT^s.
    \label{39}
\end{equation}
Using Eqs.~(\ref{37})-(\ref{38}), one has for the interfacial
temperature $T^s$
\begin{equation}
   T^s = \frac{2T^+T^-}{T^+ + T^-} = \frac{2(T_0 + L^s/c_p
   + \Delta T)(T_0 + L^s/c_p)}{2T_0 + 2L^s/c_p + \Delta T}. \label{40}\\
\end{equation}
With the accuracy of second order by $\Delta T$, this expression
gives
\begin{equation}\label{41}
    T^s = T_0 + \frac{L^s}{c_p} + \frac{\Delta T}{2}  + \cdots
\end{equation}
Taking into account Eqs.~(\ref{39}) and (\ref{41}), from
Eq.~(\ref{36}) one obtains finally the expression for the
temperature jump
\begin{equation}\label{42}
\Delta T = \frac{Rc_pV(T_0 + L^s/c_p)}{1 - RL_3/2}\Bigl \{1 -
\frac{L_3}{c_pV}\Bigl(\frac{T_M}{T_0 + L^s/c_p} - 1\Bigr)\Bigr \}\,.
\end{equation}
This expression generalizes the expression for the temperature jump
found in Ref.~\cite{PWD12} and reduces to it when $L_3 = 0$:
\begin{equation}\label{43}
 \Delta T = Rc_pV\Bigl(T_0 + \frac{L^s}{c_p}\Bigr)\,.
\end{equation}
Together with Eqs.~(\ref{37})-(\ref{38}), Eq.~(\ref{42}) allows us to obtain
equation for the interface velocity $V$. From the boundary condition~(\ref{33})
and Eqs.~(\ref{37})-(\ref{38}), one gets
\begin{equation}\label{44}
    V = - L_1\Bigl(T_M - T_0 - \frac{L^s}{c_p}  \Bigr)  +
\Bigl(\frac{L_1}{2} - L_2\Bigr)\Delta T.
\end{equation}
With the zero temperature jump, $\Delta T = 0$, one can obtain known
expression for the interface velocity which is linearly dependent
from undercooling, $\theta = T_M  - T_0 - L^s/c_p$,  with  $L_1 =
L_{00}^{'}/T_M^2 < 0$ having  the meaning of the kinetic coefficient
of crystal growth [see also Eq.~(\ref{B4})].

To simplify the further analysis, we shall use Eq.~(\ref{43}) for
the temperature jump, i.e., assume that $L_3 = 0$. Then
substituting  expression ~(\ref{43}) into Eq.~(\ref{44}), one
obtains for the interface velocity
\begin{equation}\label{45}
  V = \frac{|L_1|(T_M - T_0 - L^s/c_p)}{1 + Rc_p(|L_1|/2) + L_2  )(T_0 +
  L^s/c_p)}\,.
\end{equation}
Under the assumed approximations, Eqs.~(\ref{37})-(\ref{38}),
(\ref{43}) and (\ref{45}) completely solve  the problem of finding
boundary temperatures  and the interface velocity.

Relation~(\ref{45}) allows us to analyze the interface velocity
depending on  the initial undercooling liquid. With this aim we
rewrite Eq.~(\ref{45})
 in the dimensionless form
\begin{equation}\label{46}
  \widetilde{V} = \frac{\Delta - 1}{1 - \frac{1}{2}\widetilde{R}(1 + 2L_2/|L_1| )(\Delta - 1 -
  \widetilde{T}_M)}\,,
\end{equation}
where  $\widetilde{V} = c_pV/L^s|L_1|$ is the dimensionless
interface velocity, $\widetilde{R} = RL^s|L_1|$ is the dimensionless
Kapitza resistance, $\widetilde{T}_M = c_pT_M/L^s$ and $\Delta =
(c_p/L^s)(T_M - T_0)$ is the dimensionless undercooling. Due to inequality
$\Delta - 1 - \widetilde{T}_M = -(c_p/L^s)(T_0 + L^s/c_p) < 0$, one
gets $\Delta  < 1 + \widetilde{T}_M $.

\begin{figure}[t]\centering
    \includegraphics[width= 0.45\textwidth]{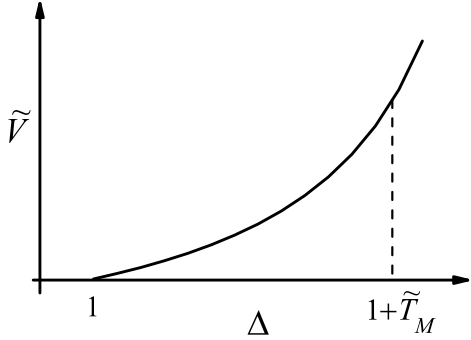}
      \parbox[t]{\textwidth}{\caption{ \label{fig1}\small Dependence of dimensionless
  velocity $\widetilde{V}$ on undercooling $\Delta$;
  $a = \frac{1}{2}\widetilde{R}(1 + 2L_2/|L_1| ) > 0$.}}
      \end{figure}

Various regimes of steady interface motion can be analyzed by Eq.~(\ref{46})
depending on $\Delta$ and the parameter
\begin{equation}\label{47}
    a = \frac{1}{2}\widetilde{R}(1 + 2L_2/|L_1| )\,,
\end{equation}
which is proportional to the Kapitza resistance $R$. Because $L_2 =
L'_{01}/T_M^2 < 0$ [see Eq.~(\ref{B5})], the parameter $a$ may have different
signs. Qualitative behavior of $\widetilde{V}$ is illustrated in
Figs.~\ref{fig1}-\ref{fig3}.  Only branches of the curves having  a
physical meaning  are shown at the figures. Depending on the value
of the parameter $a$ one can consider  three different regimes.
\begin{figure}[t]\centering
    \includegraphics[width= 0.55\textwidth]{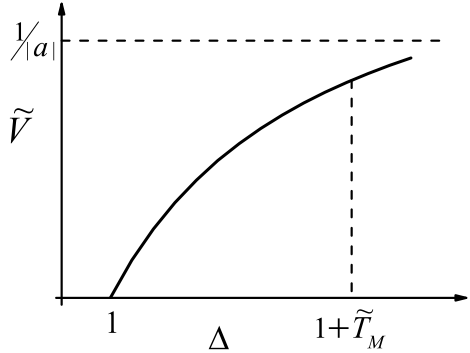}
  \parbox[t]{\textwidth}{\caption{ \label{fig2}\small Dependence of dimensionless
  velocity $\widetilde{V}$ on the undercooling  $\Delta$;
  $a = \frac{1}{2}\widetilde{R}(1 + 2L_2/|L_1| ) < 0$, $ \widetilde{T}_M +1 - 1/|a| < 1$.
  Note that, at $0 < \Delta < 1$, the velocity is negative, $\widetilde{V} < 0$. }}
    \end{figure}

 1. Figure~\ref{fig1} presents the interface velocity for $a > 0$ ($|L_1| > 2|L_2|$) within the range
 $1 < \Delta < 1 + \widetilde{T}_M$. The maximal value of the velocity is reached  at the right  border  and  is equal to
 $\widetilde{V} = \widetilde{T}_M$ (in the dimension form it is $V_{ max} = |L_1|T_M$).

\begin{figure}[t]\centering
    \includegraphics[width= 0.55\textwidth]{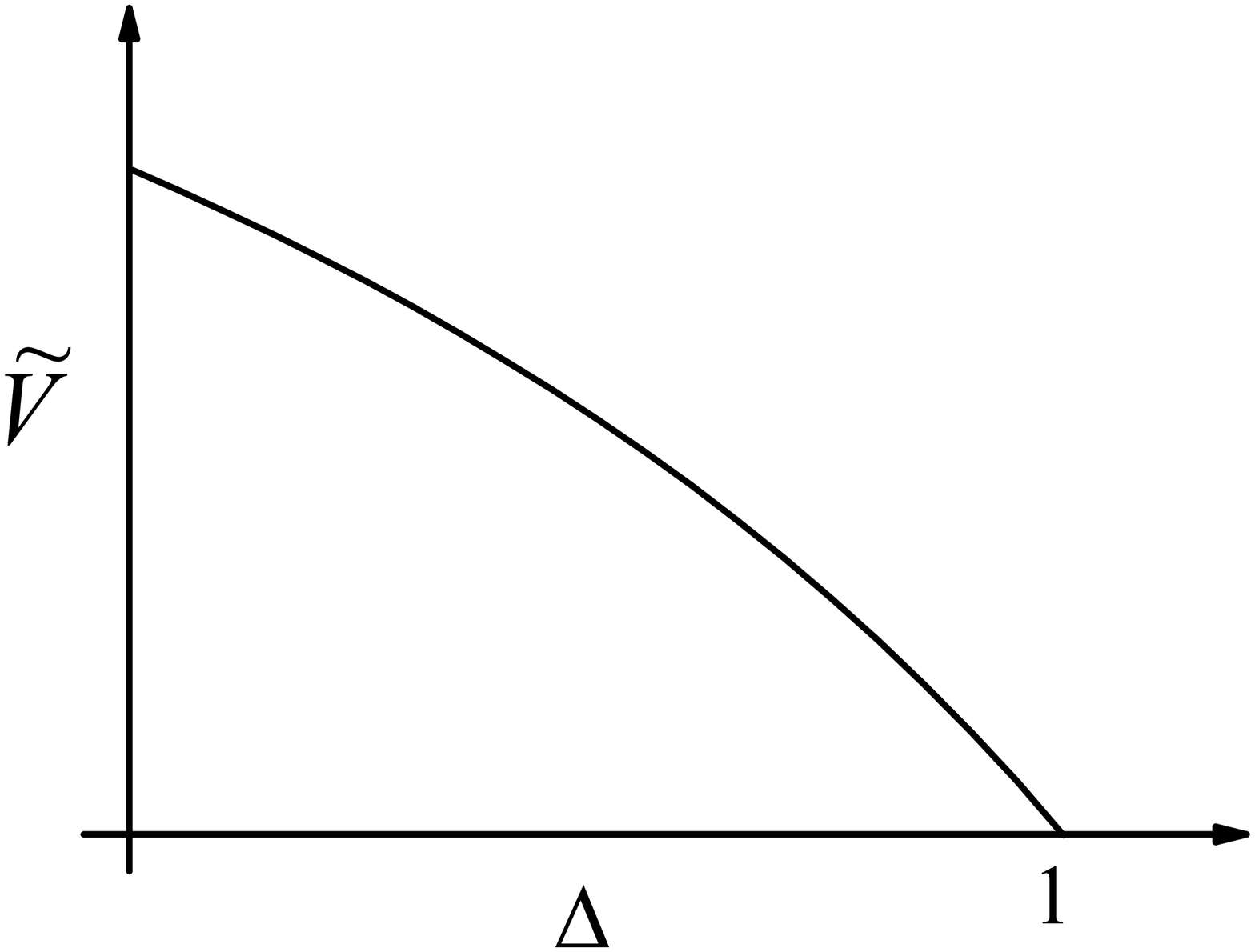}
  \parbox[t]{\textwidth}{\caption{ \label{fig3} \small Dependence of dimensionless
  velocity $\widetilde{V}$ on undercooling $\Delta$;
  $a = \frac{1}{2}\widetilde{R}(1 + 2L_2/|L_1| ) < 0$, $ \widetilde{T}_M +1 - 1/|a| > 1$.
  Note that, at $\Delta > 1$, the velocity $\widetilde{V}$
  has negative values or it reaches infinity.}}
    \end{figure}

2. Figure~\ref{fig2} presents the interface velocity for $a <0$
($|L_1| < 2|L_2|$) and $ \widetilde{T}_M +1 - 1/|a| < 1$. In the
dimension form, the latter inequality is given by
\begin{equation}\label{48}
  \frac{2}{c_pRT_M |2L_2 - L_1|} \sim  \frac{2}{c_pRT_M |L_1|} >
    1\hspace{0.5cm} (L_2 \sim L_1)\,.
\end{equation}
This case corresponds to  the small values of the Kapitza
resistance. The velocity is defined within the range of undercooling
$1 < \Delta < \widetilde{T}_M + 1$ and  at $L_2 \sim L_1$  its
maximum is of the  order of
\begin{equation}\label{49}
V_{max} \thicksim \frac{1}{|a|} \sim \frac{2}{RL^s |L_1|}.
\end{equation}

3. Figure~\ref{fig3} presents the interface velocity for $a  < 0 $
and $\widetilde{T}_M + 1 - 1/|a| > 1$ or  in the  dimensional form
\begin{equation}\label{47}
    \frac{2}{c_pRT_M |L_1|} < 1
    \hspace{0.5cm} (L_2 \sim L_1).
\end{equation}
Inequality (\ref{47}) corresponds to the large values of Kapitza
resistance.

As it follows from Fig.~\ref{fig3}, the velocity is defined within
the underrcooling range: $0 < \Delta < 1$. Due to $\Delta < 1$ and
Eq.~(\ref{38}), the temperature at the solid side of the interface
becomes larger than the equilibrium melting point: $T^- = T_0 + L^s
/c_p  > T_M$ ( the overheated boundary layer  of solid phase) but
the crystallization front moves into the liquid, $V
> 0$, as it is shown in Fig.~\ref{fig3}.
Such effect of heat trapping has been described in Refs.~\cite{U07,
UR88}. With the decrease of $T_0$, the temperature of the solid
phase  near the interface may reach the value $T_M$ ($\Delta = 1$)
with $V = 0$. In this case the  temperature at liquid side of  the
interface is also equal to the melting point $T_M$ [see
Eqs.~(\ref{37}) and (\ref{43})].

It should be noted  that  the behaviour of the velocity
 near the  value   $\Delta = 0$ in Fig.~\ref{fig3} is not physically
realized because  of the finiteness of the velocity at this point.
However the movement of the front  with  the velocity  determined by
the curve  near  the right border, i.e. close to  $\Delta = 1$, can
take place. The steady state regime of crystal growth in the range
$1 - \Delta \ll 1$  and  $\Delta < 1$ has been found in previous
works~\cite{KR98, PC90}. However, in these works,
Kapitza resistance  was not taken into account. We will return to
this issue in  next section.

As it is seen,  Figs.~\ref{fig1}-\ref{fig3} show nonlinear behavior
of the velocity against undercooling. This non-linearity appears due to
the heat Kapitza resistance. At $R = 0$, Eq.~(\ref{46}) exhibits linear
dependence $\widetilde{V} = \Delta - 1$.

\section{Concluding remarks}
\label{sec:Concl}

In the  present work boundary conditions at the solidification front
taking into account  the temperature  discontinuity at the interface
have been obtained. The interface was considered  in the framework
of Gibbs model of the dividing  surface  for which the surface
thermodynamical variables, such as  the temperature, the entropy,
the  energy, the flux densities and so on, have been introduced
independently of the corresponding bulk quantities. Boundary
conditions (\ref{7})-({\ref{10}) represent the Onsager relations at
the interface and together with Eq.~(\ref{11}) for the interface
temperature make up the complete set of equations for finding  the
boundary temperatures of the bulk phases  $T^{\pm}$, the interface
temperature $T^s$ and the solid growth velocity $\upsilon^s_n$.
Equation~(\ref{11}) describes change of the interface temperature
due to energy dissipation at the interface.

The boundary conditions (\ref{7})-({\ref{10}) have been applied to
the analysis of the steady state regime of the planar interface
motion. In the absence  of the energy dissipation at the interface
and for states close to equilibrium we have derived the expression
for the temperature jump at the interface generalizing the previous
result of Ref.~\cite{PWD12}. The obtained expression, Eq.~(\ref{42}),
is proportional to the Kapitza resistance $R$ (see
ref.~\cite{PWD12}) and it also includes the additional kinetic
coefficient $L_3$ taking into account dependence of the interfacial
energy flux from the deviation of the interface temperature $T^s$
from  equilibrium  melting temperature $T_M$ (see Eq.~(\ref{34}).

Using boundary conditions (\ref{33})-(\ref{35}), possible steady
state regimes of the crystallization front have been  analyzed
depending from the undercooling of liquid $\Delta$ and heat resistance
of Kapitza $R$. Various cases are presented in
Figs.~\ref{fig1}-\ref{fig3} for $L_3 = 0$. At $\Delta > 1$, the
steady state
motion of planar front is possible for two cases: \\
(a) if $|L_1| > 2|L_2|$, where  $L_1$ is the kinetic coefficient of
solid  growth and $L_2$ is the kinetic coefficient which defines
dependence of front velocity from temperature jump
$\Delta T$ [see Eq.~(\ref{33})]. In this case  the maximal interface velocity is equal to $V_{max} = |L_1|T_M$;\\
(b) if $|L_1| < 2|L_2|$ and at relatively small Kapitza resistance $R$,
defined by inequality~(\ref{48}). The maximal interface velocity is
given by Eq.~(\ref{49}), $V_{{max}} \sim 2/RL^s|L_1|$.

For  large enough  values of Kapitza resistance $R$ given by the
inequality~(\ref{47}), the steady state regime of crystallization is
also possible at $\Delta < 1$ in a region $1 - \Delta \ll 1$. In
such case, the interface moves with the overheated  boundary layer
of solid phase, when its temperature $T^-$  is above the equilibrium
melting temperature $T_M$, $T^-
> T_M$. The effect has been predicted in
Refs.~\cite{UR88,PC90,KR98} on the basis of phase field method. It
should be noted that  the stability of this regime should be
specially tested (see results and discussions in
Refs.~\cite{KR98,PC90}).

Finally, let us make one essential remark. The whole
analysis has been done for the constant Kapitza resistance $R$
independent from temperature. However, measurements made at low and
high temperatures~\cite{SP89,CZB12} show that the heat resistance
increases by the power law with decreasing temperature. In
solidification experiments, therefore,  the dependence of Kapitza
coefficient  from the melt initial temperature  or undercooling  is
possible that can be qualitatively expressed, for example, as
\begin{eqnarray*}
  &&  R = R_0 + R_1\Delta^\beta\hspace{1.5cm}(\beta > 0),\nonumber
\end{eqnarray*}
with positive constant $R_0$ and $R_1$. At $R_1 \ll R_0$, the
Kapitza resistance is almost independent from the temperature, $R \approx
R_0$. In this case, depending on  the relation between  the kinetic
coefficients,  the interface velocity will behave analogously to the
curves shown in Figs.~\ref{fig1} and~\ref{fig2}.

In the liquid phase with $R_1 \gtrsim R_0$, the heat resistance $R$
may take high enough values at $\Delta \thicksim 1$, $R \thicksim
R_0 + R_1$, for which the steady state regime can be realized
analogously to the regime shown in Fig.~\ref{fig3} closely to
$\Delta = 1$. If such regime is realized, then, it can be detected
indirectly. As it can be seen in Fig.~\ref{fig3}, the front velocity
decreases with the increase of undercooling $\Delta$. Quite possible, due to
the large heat resistance at high undercooling, the solidification rate slows down
as it was been observed in experiments~\cite{BHB02,CACA09, H14}.

\appendix
 \section{The entropy production}
 \label{app1}

In this Appendix, the method of Bedeaux, Albano and Mazur
(BAM-method) is applied to the problem of pure liquid
solidification. Details of mathematical formalism are given in
Refs.~\cite{BAM76,ZB82} (see also Ref.~\cite{CCR84}).

To derive entropy production~$\sigma^s$ we use conservation laws
in the following form~\cite{GM69}:
\begin{eqnarray}
  \frac{\partial\varrho\bm{\upsilon}}{\partial t}
  &=& - \,\mbox{div}\,{\bf P }, \label{A1}\\
  \frac{\partial e}{\partial t} &=& - \,\mbox{div}\,{\bf J
  }_e,\label{A2}\\
\frac{\partial s}{\partial t} &=& - \,\mbox{div}(s\bm{\upsilon} +
{\bf J }_s) +  \sigma,\label{A3}
\end{eqnarray}
where $\varrho$, $e$ and $s$ are the densities of mass, energy and
entropy, respectively, given by Eq.~(\ref{1}). Analogously to
Eq.~(\ref{1}), one can write the stress tensor $\bf P $ and energy
flux ${\bf J }_e = {\bf P }\cdot\bm{\upsilon} + e\bm{\upsilon} +
{\bf J}_q$, where ${\bf J}_q$ is the heat flux and ${\bf J}_s$ is
the entropy flux.

To obtain equations for densities $x^\pm $ and $x^s $, the
quantities of the form Eq.~(\ref{1}) are substituted in
Eqs.~(\ref{A1})--(\ref{A3}) and the following relations are used~\cite{BAM76}:
\begin{eqnarray*}
  \frac{\partial\mathit\Theta ^\pm }{\partial t} &=& \mp\, \upsilon^{s}_n\delta^s({\bf r}t) \label{subeq:1}\\
 \nabla\mathit\Theta ^\pm  &=& \pm\,{\bf n} \delta^s({\bf r}t)\,\label{subeq:2}\\
\frac{d^s}{dt}\delta^s &\equiv& \Bigl(\frac{\partial }{\partial t} +
\bm{\upsilon}^{s}\cdot\nabla\Bigr )\delta^s = 0,\label{subeq:3}
\end{eqnarray*}
where $d^s/dt$ is the total derivative at the interface,
$\upsilon^{s}_n = {\bf n}\cdot\bm\upsilon^{s}$ is the normal
components of the vector $\bm\upsilon^s $ and ${\bf n}$ is the unit
normal vector to the interface directed to the liquid. After
substitution, the coefficients of ${\mathit\Theta}^+$,
${\mathit\Theta}^-$ and ${\mathit\delta }^s$ in each equation vanish
separately ~\cite{BAM76}. If the term of the form of ${\bf
A}\cdot{\nabla\mathit\delta }^s$ appears then the normal component
of the vector ${\bf A}$ also must vanish, $A_n = {\bf n}\cdot{\bf A}
= 0$.

As a result, Eq.~(\ref{A1}) gives (with $ \varrho\bm\upsilon = 0$):
\begin{equation}\label{A4}
{\mathit\Theta }^+\mbox{div}\cdot{\bf P }^+ + {\mathit\Theta
}^-\mbox{div}\cdot{\bf P }^- + \bigl[ {\bf n}\cdot({\bf P }^+ - {\bf
P }^-) + \mbox{div}\cdot{\bf P }^s \bigr]{\mathit\delta }^s +
{\nabla\mathit\delta }^s\cdot{\bf P }^s = 0\,.
\end{equation}
From this it follows
\begin{eqnarray}
 && \mbox{div}\cdot{\bf P }^\pm  = 0,    \label{A5}\\
  &&\mbox{div}\cdot{\bf P }^s + {\bf n}\cdot({\bf P }^+ - {\bf P
}^-)   = 0, \label{A6}\\
& & {\bf n}\cdot{\bf P }^s = 0. \label{A7}
\end{eqnarray}
Equation~(\ref{A5}) is the usual equation of equilibrium in each
phase and expressions (\ref{A6}) and (\ref{A7}) represent boundary
conditions at the interface. In Eq.~(\ref{A6}),  ${\bf P }^\pm$ are
taken at the surface (due to surface $\delta$-function in
Eq.~(\ref{A4})), i.e., are  the limiting values of the stress tensor
at corresponding side of the interface. It what follows,  if the bulk
quantities $x^\pm$ occur in equations related to the surface (in
a manner of Eq.~(\ref{A4})), their limiting values at the surface
are implied.

The similar treatment of Eq.~(\ref{A2}) gives for the surface
density~$e^s$:
\begin{eqnarray}\label{A8}
  &&  \frac{\partial e^s}{\partial t} = - \,\mbox{div}\cdot({\bf P
    }^s\cdot\bm{\upsilon}^s + e^s\bm{\upsilon}^s + {\bf J}^s_q)
  + [e\upsilon^{s}_n -
    J_{qn}]_-,  \\
    && J^s_{qn} = {\bf n}\cdot{\bf J}^s_{q} = 0\label{A9}\,,
\end{eqnarray}
where for every bulk quantity $x^\pm$ taken at the interface we introduce the notations
\[[x]_- = x^+ - x^{-}\]
and $J^{\pm}_{qn} = {\bf n}\cdot{\bf J}^{\pm}_{q}$.  From
Eq.~(\ref{A9}) follows that the interface heat flux is directed only
along the surface.

Since $\varrho^s = 0$, the total interfacial energy $e^s$ is equal
to the internal  energy $u^s$
\[e^s = u^s + \frac{1}{2}\varrho^s(\bm{\upsilon}^s)^2 = u^s\,.\]
Therefore Eq.~(\ref{A8}) also presents the equation for $u^s$. The
equation for  total derivative of $u^s$ follows from Eq.~(\ref{A8})
\begin{eqnarray}
  &&  \frac{d^su^s}{dt} = - \,\mbox{div}({\bf P
    }^s\cdot\bm{\upsilon}^s +  {\bf J}^s_q) - e^s\mbox{div}\,\bm{\upsilon}^s
 + [e\upsilon^{s}_n -
J_{qn}]_-.  \label{A10}
\end{eqnarray}
The  equations  in  the bulk give usual the energy conservation laws
in continuum medium.

Using the similar method, equation for the entropy balance gives for
the entropy density $s^s$ at the interface
\begin{eqnarray}
   & & \frac{d^ss^s}{dt} = -  \,\mbox{div}\,{\bf J^s_s} -s^s\mbox{div}\,\bm{\upsilon}^s
   + [s\upsilon^{s}_n -
    J_{sn}]_- + \sigma^s,   \label{A11}  \nonumber\\
  & & J_{sn}^s = {\bf n}\cdot{\bf J}^s_{s} = 0,
\end{eqnarray}
where ${\bf J}^s_{s}$ is the surface entropy flux   and $J_{sn}^\pm
= {\bf n}\cdot{\bf J}_s^\pm$ is the normal projection of the entropy
flux ${\bf J}_s^\pm$ from the bulk phases such that ${\bf J}_s^\pm =
{\bf J}_q^\pm/T^\pm$~\cite{GM69}.

Taking into account the equality~(\ref{5}), we assume now that the
following   equality is fulfilled on the moving surface,
\begin{equation}\label{A12}
\frac{d^ss^s}{dt} = \frac{1}{T^s}\frac{d^su^s}{dt}\,.
\end{equation}
After substituting Eq.~(\ref{A10}) into Eq.~(\ref{A12}), one gets
\begin{eqnarray}
\frac{d^ss^s}{dt}  &=& - \,\mbox{div}\,\frac{{\bf
 J}^s_q}{T^s} + {\bf
 J}^s_q\cdot\nabla\frac{1}{T^s}
- \frac{1}{T^s}{\bf \Pi}^s:\nabla\bm{\upsilon}^s  -
\frac{p^s}{T^s}\,\mbox{div}\,\bm{\upsilon}^s -
\nonumber\\
   && -
\frac{1}{T^s}e^s\,\mbox{div}\,\bm{\upsilon}^s
   + \frac{1}{T^s}[{\bf n}\cdot{\bf
P}\cdot\bm{\upsilon}^s + e{\upsilon}^s_n -
   J_{qn}]_-\,.
  \label{A13}
\end{eqnarray}
Here the stress tensor ${\bf P}^s$ is divided into the viscous part
${\bf \Pi}^s$ and elastic part $p^s$, such that ${\bf P}^s = {\bf
\Pi}^s + p^s(1 - {\bf n}{\bf n})$, where $p^s$ has the meaning of
surface tension (with the accuracy to the sign)~\cite{BAM76}. We also
assume that the medium is isotropic and $(1 - {\bf n}{\bf n})$ is
the tensor with the following components~$\delta_{\alpha\beta} -
n_\alpha n_\beta$, and $n_\alpha$ is the projections of the vector
$\bf n$. The tensor $(1 - {\bf n}{\bf n})$ projects vectorial and
tensorial quantities to the tangent plane to the interface and
provides the condition ${\bf \Pi}^s\cdot \bf n = 0$ which is follow
from Eq.~(\ref{A7}).

After some algebra, comparison of Eq.~(\ref{A13}) with Eq.~(\ref{A11}) gives
\begin{equation}
{\bf J}^s_s  = \frac{{\bf J}^s_q}{T^s} \label{A14}
\end{equation}
\begin{eqnarray}
 &&\sigma^s = - \frac{1}{T^s} \stackrel{\circ}{{\bf\Pi}^s}:\nabla\bm{\upsilon}^s -
 {\bf J}^s_q \cdot\Big(\frac{1}{T^s}\Big)^2\nabla T^s   +
 \frac{1}{T^s}\Bigl[({\bf n\cdot{\bf \Pi}})_{||}\Bigr]_-\cdot\bm{\upsilon}^s_{||} -   \nonumber  \\
  && - \frac{\pi^s}{T^s} \mbox{div}\,\bm{\upsilon}^s  + \frac{1}{T^s}\Bigl[\mu\varrho +
  \Pi_{nn}\Bigr]_-\upsilon^s_{n} + [J_{qn} - sT\upsilon^s_{n}]_+ \Bigl(\frac{1}{T^+} - \frac{1}{T^-}
  \Bigr) + \nonumber  \\
  &&\phantom{aaaaaaaaa} + [J_{qn} - sT\upsilon^s_{n}]_{-} \Bigl[ \frac{1}{2}\Bigl( \frac{1}{T^+} + \frac{1}{T^-}\Bigr) -
  \frac{1}{T^s}\Bigr]\,,\label{A15}
\end{eqnarray}
here $[x]_+ = \frac{1}{2}(x^+ + x^-)$ for $x^{\pm}$ at the interface,
${\bf \Pi}^{\pm} = {\bf P}^{\pm} - p^{\pm}1$ defines the viscous
part of stress tensor ${\bf P}^{\pm}$ with $p^{\pm}$ being the
hydrostatic pressure, and
\begin{eqnarray*}
&& \stackrel{\circ}{{\bf\Pi}^s} = {\bf\Pi}^s - \pi^s(1 - {\bf n}{\bf
n}) \\
  &&\pi^s = \frac{1}{2}\mbox{Sp}{\bf\Pi}^s\\
  &&({\bf n\cdot{\bf \Pi}})_{||} = (1- {\bf n}{\bf n})\cdot({\bf n\cdot{\bf
  \Pi}})\\
  && \Pi_{nn} = {\bf n}\cdot {\bf \Pi}\cdot{\bf n}\,,
\end{eqnarray*}
where $\bm{\upsilon}^s_{||}$ is the   component  of
$\bm{\upsilon}^s$ tangent to the interface and $\mu^{\pm}$ are
chemical potentials in bulk phases. In the derivation of
Eq.~(\ref{A15}) we used the relations
\begin{equation}\label{A16}
\mu^{\pm}\varrho^{\pm} = e^{\pm} - T^{\pm}s^{\pm} +  p^{\pm}
\end{equation}
for the bulk quantities \cite{GM69} and
\begin{equation}\label{A17}
e^{s} =  T^{s}s^s -  p^{s}
\end{equation}
for the surface energy density with $\varrho^s = 0$ \cite{BAM76}.

Relations~(\ref{A15}) define the surface entropy production taking
into account the heat transfer along the interface as well as
effects of the surface and the bulk viscosities (see terms in
Eq.~(\ref{A15}) containing viscous stress tensors
$\stackrel{\circ}{{\bf\Pi}^s}$, $\pi^s, {\bf \Pi}^\pm$ ).  The
relation ~(\ref{A15}) is greatly simplified  if we  neglect  the
surface viscosity and take into account  that in the stagnant
medium, ${\bf \Pi}^\pm = 0$. In such case the interfacial entropy
production takes the following form
\begin{eqnarray}
 &&\sigma^s =  -
 {\bf J}^s_q \cdot\Big(\frac{1}{T^s}\Big)^2\nabla T^s
 +  \frac{1}{T^s}\Bigl[\mu\varrho \Bigr]_-\upsilon^s_{n} +
\nonumber  \\
  && \phantom{aaaa} + [J_{qn} - sT\upsilon^s_{n}]_+ \Bigl(\frac{1}{T^+} - \frac{1}{T^-}
  \Bigr) +\label{A18} \\
  &&\phantom{aaaa}+ [J_{qn} - sT\upsilon^s_{n}]_{-} \Bigl[ \frac{1}{2}\Bigl( \frac{1}{T^+} + \frac{1}{T^-}\Bigr) -
  \frac{1}{T^s}\Bigr]\,.\nonumber
\end{eqnarray}
Introducing now the bulk transverse flux of energy, $J_E = J_{qn}
- (sT + \mu\varrho)\upsilon^s_{n} =  J_{qn} - h\upsilon^s_{n}$,
where $h = sT + \mu\varrho = e + p$ is the enthalpy, the entropy
production Eq.~(\ref{A18}) can be finally rewritten as
\begin{eqnarray}
 &&\sigma^s =  -
 {\bf J}^s_q \Bigl(\frac{1}{T^s}\Big)^2\nabla_{||} T^s
 + \Bigl[\frac{\mu\varrho }{T}\Bigr]_-\upsilon^s_{n} + \nonumber  \\
  &&\phantom{aaaa} + [J_{qn} - h\upsilon^s_{n}]_+ \Bigl(\frac{1}{T^+} - \frac{1}{T^-}
  \Bigr) +\label{A19} \\
  &&\phantom{aaaa}+ [J_{qn} - h\upsilon^s_{n}]_{-} \Bigl[ \frac{1}{2}\Bigl( \frac{1}{T^+} + \frac{1}{T^-}\Bigr) -
  \frac{1}{T^s}\Bigr]\,.\nonumber
\end{eqnarray}
In obtaining Eq.~(\ref{A19}), there was taken into account that the
surface heat flux lies in the tangent plane and it has zero scalar
product with any vector being normal to the surface. The symbol $||$ means
the component of a vector along tangent plane and  $\nabla_{||} = (1
- {\bf n}{\bf n})\cdot\nabla $.

Equality~(\ref{A19}) should be added by the equation for the
interfacial temperature which can be derived from the equation for
the interfacial internal energy~(\ref{A10}). Assuming that $u^s =
u^s(T^s)$,
\[\frac{d^su^s}{dt} = c^s\frac{d^sT^s}{dt}\,,\]
where $c^s$ is the heat capacity of the interface, and using equalities
(\ref{A16}) and ${\bf P}^s = {\bf \Pi}^s + p^s(1 - {\bf n}{\bf n})$,
one obtains from Eq.~(\ref{A10}) the following equation for interfacial temperature:
\begin{eqnarray}
&&c^s\frac{d^sT^s}{dt} = -  \,\mbox{div}\, {\bf J}^s_q - {\bf
\Pi}^s:\nabla\bm{\upsilon}^s + [{\bf n}\cdot{\bf
\Pi}\cdot\bm{\upsilon}^s]_- \nonumber\\
&&\phantom{c^s\frac{d^sT^s}{dt} =}-
T^ss^s\mbox{div}\,\bm{\upsilon}^s + [h\upsilon^{s}_n -
    J_{qn}]_-  \,. \label{A20}
\end{eqnarray}
 Neglecting the viscosity, one obtains finally
\begin{equation}\label{A21}
c^s\frac{d^sT^s}{dt} +  \,\mbox{div}\, {\bf J}^s_q  +
T^ss^s\mbox{div}\,\bm{\upsilon}^s  =  [h\upsilon^{s}_n -
    J_{qn}]_-\,.
\end{equation}

\section{Interface kinetics}\label{app2}
In the absence of local equilibrium at the solidification front the
usual kinetic equation for the rate of phase transformations
\cite{Ch84} must be modified to take into account for the
temperature discontinuity across the interface.  In this section we
consider a possible modification that is appropriate for the model
of Gibbs, when the interface has its own temperature different from
the boundary temperature of the bulk phases.

During crystallization the atoms must overcome an energy barrier
separating different phases. Energy spatial distribution, shown in
Fig.~\ref{fB1} in which $Q^\pm$ are the heights of energy barriers,
is used for both continuous and discontinuous changes of temperature
across the interface~\cite{Ch84,GD00}. If the surface phase is
characterized by the temperature $T^s$ (which is different from the
temperature $T^+$ and $T^-$ from sides of the interface), one can
suppose to use energy profile as depicted in Fig. 5.
\begin{figure}[t]\centering
    \includegraphics[width=0.36\textwidth]{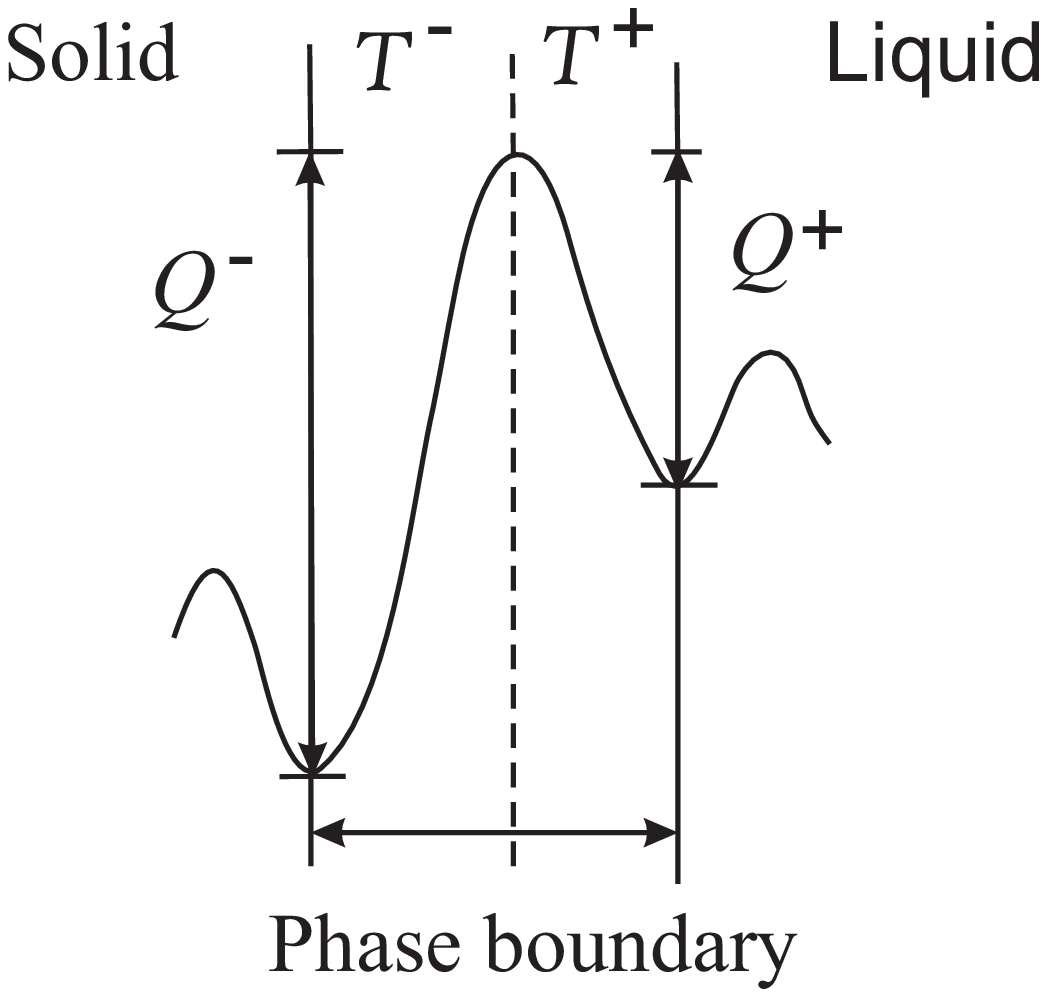}
   \hspace{1.8cm}
  \includegraphics[width=0.45\textwidth]{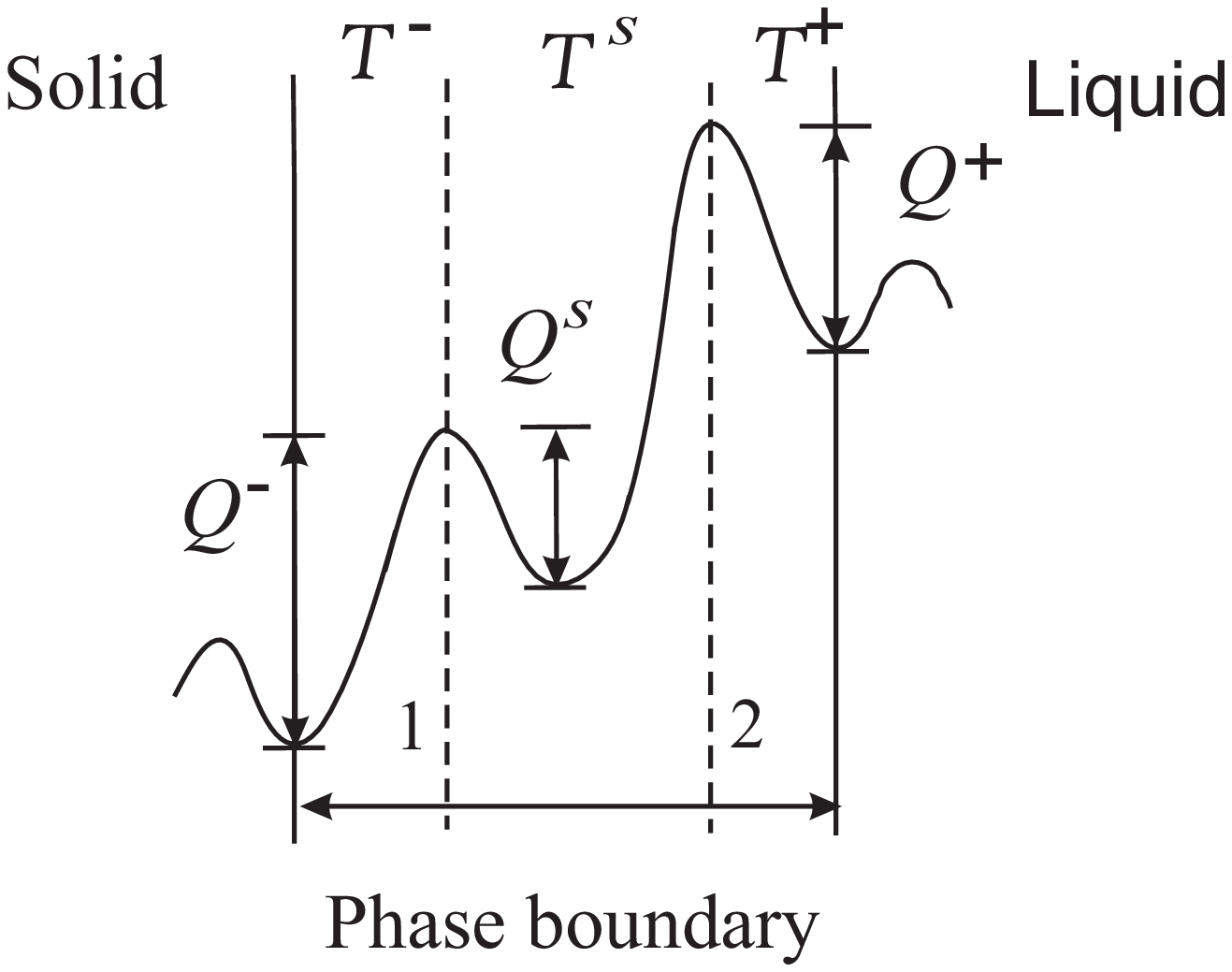}
\\
\parbox[t]{0.46\textwidth}{\caption { \label{fB1}\small The atomic potential profile through the phase boundary
with the  temperature jump $T^+ - T^-$.}}
 \hspace{0.7cm}
 \parbox[t]{0.46\textwidth}{\caption{ \label{fB2}\small The atomic potential profile through the
 phase boundary, that is considered as a thermodynamic system with its own temperature
 $T^s$.}}
\end{figure}
Atoms which overcome the barrier between liquid and surface layer 2
at the temperature $T^+$ should also overcome the barrier $Q^s$
between surface 1 and solid phase at the temperature $T^s$.

Thus, the process of transition of particles from the liquid phase
in the solid can be described as follows. If $j^+ =
A_2e^{-Q^+/kT^+}$ (with $k$ being Boltzmann's constant) is the flux
for the atomic attachment to the boundary 1, then only a part of
this flux, $A_sj^+e^{-Q^s/kT^s}$, may reach the solid phase.
The atomic detachment flux from the solid phase is given by $j^- =
Be^{-Q^-/kT^-}$. For simplicity, we assume that the return flux from
the interface into the liquid phase (through boundary 2) is
negligible, i.e. the energy barrier in this direction is large
enough. In this case, the rate of the growth of the  new phase
should mainly be determined by kinetics on the boundary 1. Then, the
flux difference at the boundary 1 gives the equation for the
growth velocity
\begin{equation}\label{B1}
  \upsilon^s = Ae^{-Q^+/kT^+}e^{-Q^s/kT^s} - Be^{-Q^-/kT^-}\,,
\end{equation}
where $A=A_2A_s$ and $B$ are positive constants having the dimension of
velocity. In equilibrium, one has $\upsilon^s=0$ at the melting
temperature $T_M$  that allous us to write down   Eq.~(\ref{B1}) as
\begin{eqnarray}\label{B2}
\upsilon^s =  Ae^{- Q^+/kT^+ - Q^s/kT^s}\times\phantom{aaaaaaaaaaaaaaaaaaaaaaaaaaaaaaaa}&&\nonumber\\
 \Biggl\{1 - \exp\Biggl[\frac{[Q]_- + Q^s}{k}\Biggl(\frac{1}{T^s} - \frac{1}{T_M}\Biggr)\Biggr]
 \exp\Biggl[\frac{[Q]_+}{k}\Biggl(\frac{1}{T^+} - \frac{1}{T^-}\Biggr)\Biggr]\times &&\nonumber\\
 \times \exp\Biggl[- \frac{[Q]_-}{2k}\Biggl( \frac{2}{T^s} -  \frac{1}{T^+} - \frac{1}{T^-}\Biggr)\Biggr]
  \Biggr\}\,,\phantom{aaaaaaaaaa}
  &&\nonumber\\
\end{eqnarray}
where $[Q]_- = Q^+ - Q^-$, $[Q]_+ = \frac{1}{2}(Q^+ + Q^-)$.

Due to Eq.~(\ref{14}),
\[\frac{1}{T^s} =  \frac{1}{2}\Biggl(\frac{1}{T^+} + \frac{1}{T^-}\Biggr)\,,\]
 one can gets from Eq.~(\ref{B2})
\begin{eqnarray}
&&  \upsilon^s =  Ae^{- Q^+/kT^+} \exp \Biggl[-
\frac{Q^s}{2k}\biggl(\frac{1}{T^+}
  +
  \frac{1}{T^-}\Bigr)\Biggr]
 \Biggl\{1 - \exp\Biggl[\frac{[Q]_+}{k}\Biggl(\frac{1}{T^+} -
\frac{1}{T^-}\Biggr)\Biggr]\times\nonumber\\
&&\phantom{aaaaaaaa} \times\exp\Biggl[\frac{[Q]_- +
Q^s}{2k}\Biggl(\frac{1}{T^+} + \frac{1}{T^-} - \frac{2}{T_M}
  \Biggl)
  \Biggl]
  \Biggr\}\label{B3}
\end{eqnarray}
It is easy to show that when the temperature
continuously changes across the interface, $T^+ = T^-$, and there is no
internal energy barrier, $Q^s = 0$, Eq.~(\ref{B3}) is reduced
to the known expression for the solidification rate of a pure liquid
with $Q^- - Q^+$ being latent heat solidification~\cite{Ch84}.
In the case of the temperature discontinuity at the
phase boundary, $T^+ \neq T^- $, and at $Q^s=0$, Eq.~(\ref{B3}) coincides
with the kinetic equation used in the work~\cite{GD00}.

The equality~(\ref{B3}) allows us to obtain kinetic coefficients
$L'_{00}$  and $L'_{01}$ from Eq.~(\ref{16}) by the energy
parameters. Indeed, expanding Eq.~(\ref{B3}) in powers $1/T^s - 1/T_M$ and
$1/T^+ - 1/T^-$, and comparing the result with Eq.~(\ref{16}), one obtains
for the kinetic coefficients
\begin{eqnarray}
  L_{00}' &=& -\frac{A}{k}(Q^s + [Q]_-)e^{- (Q^+ + Q^s) /kT_M}, \label{B4}\\
  L_{01}' &=& -\frac{A}{k} [Q]_+ e^{- (Q^+ + Q^s)/kT_M} < 0\,.
 \label{B5}
\end{eqnarray}

\section*{Acknowledgements}\label{ack}
P.\,K.\,G. acknowledges the support from German Research Foundation (DFG Project RE
1261/8-2) and from the Russian Science Foundation (project 16-11-10095).








\newpage



\end{document}